\documentclass[pra,twocolumn,amsmath,amssymb,aps,showpacs]{revtex4}
\usepackage{graphicx}
\usepackage[utf8]{inputenc}
\usepackage{bm}
\usepackage{subfigure}
\usepackage{amsmath}
\newcommand{\pr}{Phys. Rev.\ }

\newcommand{\jpa}{J. Phys. A\ }

\newcommand{\etal}{{\em et al. }}
\newcommand{\etals}{{\em et al.}}

\newcommand{\UQ}{School of Mathematics and Physics, University of Queensland, Brisbane, 
Queensland 4072, Australia.}

\begin{document}

\title{Entanglement and asymmetric steering over two octaves of frequency difference with cascaded harmonic generation}

\author{M.~K. Olsen}
\affiliation{\UQ}
\date{\today}

\begin{abstract}

We analyse a nonlinear optical system which uses cascaded nonlinearities to produce both second and fourth harmonic outputs from an input field at the fundamental frequency. Using fully quantum equations of motion, we show that the system produces quadrature squeezed outputs which exhibit bipartite entanglement, EPR-steering, and asymmetric steering across a two octave frequency range.

\end{abstract}

\pacs{42.50.Dv,42.65.Ky,03.65.Ud,03.67.Bg}  

\maketitle


The theory of the interaction of light fields at one frequency with nonlinear materials to produce fields at different frequencies goes back at least to Armstrong \etals, who produced a seminal work which included second and third harmonic generation~\cite{Armstrong}. That work did not consider fourth harmonic generation, possibly because the nonlinearity needed for a five wave mixing process would be relatively weak. Despite this inherent weakness, and the difficulty of finding materials that are transparent over two octaves, Komatsu \etal successfully produced fourth harmonic from Li$_{2}$B$_{4}$O$_{7}$ crystal, with a conversion efficiency of $20\%$~\cite{Kumatsu}. The advent of quasi-periodic superlattices meant that higher than second order processes were readily available, with Zhu \etal producing third harmonic by coupling second harmonic (SHG) and sum-frequency generation in 1997~\cite{Zhu}. Using CsLiB$_{6}$O$_{10}$, Kojima \etal were able to produce fourth harmonic at a 10 kHz repetition rate by 2000~\cite{Kojima}. Broderick \etal have produced fourth harmonic from a cascaded SHG process using a  HeXLN crystal~\cite{Broderick}, which could  be tuned for both processes at the same temperature. Südmeyer \etal produced fields at both second and fourth harmonics using an intracavity cascaded process with LBO and BBO crystals, with greater than $50\%$ efficiency in 2007~\cite{Sudmeyer}. More recently, Ji \etal have generated light at 263 nm from a 1053 nm input, using KD$^{*}$P and NH$_{4}$H$_{2}$PO$_{4}$ crystals with non-critical phase matching~\cite{Ji}.
In this work, we combine these technical and experimental advances with quantum information techniques and propose a useful quantum technology spanning two octaves. 

The theoretical examination of the quantum statistical properties of fourth harmonic generation began with Kheruntsyan \etals, who analysed an intracavity cascaded frequency doubler process~\cite{Yerevan}. The authors adiabatically eliminated the highest frequency mode to calculate squeezing in the remaining modes, while also finding self-pulsing in the intensities. Yu and Wang~\cite{Yu} performed an analysis of the system without any elimination, starting with the full positive-P representation~\cite{P+} equations of motion. Linearising around the steady-state solutions of the semi-classical equations, they performed a stability analysis and examined the entanglement properties using the method of symplectic eigenvalues~\cite {Serafini}. 

In this work we extend previous analyses by integrating the full positive-P equations in both the travelling wave and intracavity configurations. We  
show that all three fields can exhibit quadrature squeezing in both configurations and that the full quantum solutions for the field intensities can be qualitatively different from the classical predictions in both the travelling wave and intracavity situations. We also use the Reid EPR criteria~\cite{EPRMDR} to detect bipartite entanglement and EPR steering~\cite{EPR,Erwin,Wisesteer} in all three possible bipartitions. We find that as well as producing steering and entanglement across one octave for both transitions, the system can also be used to produce entangled states and asymmetric steering across two octaves of frequency difference.


Our system consists of three fields interacting in nonlinear materia, which could either be a periodically poled dielectric or two separate nonlinear crystals held in the same optical cavity. The equations of motion are the same for both.The fundamental field at $\omega_{1}$, which will be externally pumped, is represented by $\hat{a}_{1}$. The second harmonic, at $\omega_{2}=2\omega_{1}$, is represented by $\hat{a}_{2}$, and the fourth harmonic, at $\omega_{3}=4\omega_{1}$, is represented by $\hat{a}_{3}$. The nonlinearity $\kappa_{1}$ couples the fields at $\omega_{1}$ and $\omega_{2}$, while $\kappa_{2}$ couples those at $\omega_{2}$ and $\omega_{3}$. The unitary interaction Hamiltonian in a rotating frame is then written as 
\begin{equation}
{\cal H}_{int} = \frac{i\hbar}{2}\left[ \kappa_{1}(\hat{a}_{1}^{2}\hat{a}_{2}^{\dag}-\hat{a}_{1}^{\dag\,2}\hat{a}_{2})+\kappa_{2}(\hat{a}_{2}^{2}\hat{a}_{3}^{\dag}-\hat{a}_{2}^{\dag\,2}\hat{a}_{3}) \right].
\label{eq:UHam}
\end{equation}
For the intracavity configuration, we also have the pumping Hamiltonian,
\begin{equation}
{\cal H}_{pump} = i\hbar\left(\epsilon\hat{a}_{1}^{\dag}-\epsilon^{\ast}\hat{a}_{1} \right),
\label{eq:Hpump}
\end{equation}
where $\epsilon$ represents an external pumping field which is usually taken as coherent, although this is not necessary~\cite{Liz}. The damping of the cavity into a zero temperature Markovian reservoir is described by the Lindblad superoperator 
\begin{equation}
{\cal L}\rho = \sum_{i=1}^{3}\gamma_{i}\left(2\hat{a}_{i}\rho\hat{a}_{i}^{\dag}-\hat{a}_{i}^{\dag}\hat{a}_{i}\rho-\rho\hat{a}_{i}^{\dag}\hat{a}_{i} \right),
\label{eq:Lindblad}
\end{equation}
where $\rho$ is the system density matrix and $\gamma_{i}$ is the cavity loss rate at $\omega_{i}$. In this work we will treat all three optical fields as being at resonance with the optical cavity.

Following the usual procedures~\cite{DFW,QNoise}, we can derive equations of motion in the positive-P representation~\cite{P+},
\begin{eqnarray}
\frac{d\alpha_{1}}{dt} &=& \epsilon-\gamma_{1}\alpha_{1}+\kappa_{1}\alpha_{1}^{+}\alpha_{2}+\sqrt{\kappa_{1}\alpha_{2}}\,\eta_{1}, \nonumber \\
\frac{d\alpha_{1}^{+}}{dt} &=& \epsilon-\gamma_{1}^{+}\alpha_{1}^{+}+\kappa_{1}\alpha_{1}\alpha_{2}^{+}+\sqrt{\kappa_{1}\alpha_{2}^{+}}\,\eta_{2}, \nonumber \\
\frac{d\alpha_{2}}{dt} &=& -\gamma_{2}\alpha_{2}+\kappa_{2}\alpha_{2}^{+}\alpha_{3}-\frac{\kappa_{1}}{2}\alpha_{1}^{2}+\sqrt{\kappa_{2}\alpha_{3}}\,\eta_{3}, \nonumber \\
\frac{d\alpha_{2}^{+}}{dt} &=& -\gamma_{2}\alpha_{2}^{+}+\kappa_{2}\alpha_{2}\alpha_{3}^{+}-\frac{\kappa_{1}}{2}\alpha_{1}^{+\,2}+\sqrt{\kappa_{2}\alpha_{3}^{+}}\,\eta_{4}, \nonumber \\
\frac{d\alpha_{3}}{dt} &=& -\gamma_{3}\alpha_{3}-\frac{\kappa_{2}}{2}\alpha_{2}^{2}, \nonumber \\
\frac{d\alpha_{3}^{+}}{dt} &=& -\gamma_{3}\alpha_{3}^{+}-\frac{\kappa_{2}}{2}\alpha_{2}^{+\,2},
\label{eq:Pplus}
\end{eqnarray}
noting that these have the same form in either It\^o or Stratonovich calculus~\cite{SMCrispin}.
The complex variable pairs $(\alpha_{i},\alpha_{j}^{+})$ correspond to the operator pairs $(\hat{a}_{i},\hat{a}_{j}^{\dag})$ in the sense that stochastic averages of products converge to normally-ordered operator expectation values, e.g. $\overline{\alpha_{i}^{+\,m}\alpha_{j}^{n}} \rightarrow \langle \hat{a}_{i}^{\dag\,m}\hat{a}_{j}^{n} \rangle$. The $\eta_{j}$ are Gaussian noise terms with the properties $\overline{\eta_{i}}=0$ and $\overline{\eta_{j}(t)\eta_{k}(t')}=\delta_{jk}\delta(t-t')$. We note here that our equations have a sign change from those used by Yu and Wang~\cite{Yu}, and that this has no physical significance.

\begin{figure}[tbhp]
\includegraphics[width=0.75\columnwidth]{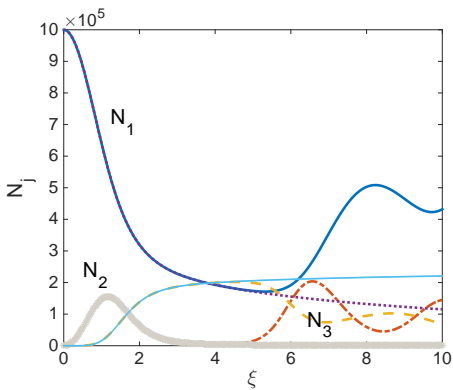}
\caption{(colour online) Positive-P and classical solutions for intensities in the travelling wave configuration, with $\kappa_{1}=.005$, $\kappa_{2}=4\kappa_{1}$, and $N_{1}(0)=10^{6}$. The quantum solutions are the lines which begin oscillations at $\xi\approx 5$. The dimensionless interaction time, $\xi$, is equal to $\kappa_{1} |\langle\alpha_{1}(0)\rangle |t$. The equations were averaged over $1.4\times 10^{6}$ stochastic trajectories. All quantities in this and subsequent figures are dimensionless.}
\label{fig:Ntrav}
\end{figure}


Before we analyse the properties of the intracavity system, it is of interest to look at a simplified travelling wave configuration, using only the unitary Hamiltonian dynamics. The equations for this are found by removing the pumping and damping terms from Eq.~\ref{eq:Pplus}. The time development of the intensities is shown in Fig.~\ref{fig:Ntrav}, both in the fully quantum picture and in a semi-classical approximation where the noise terms are removed from the equations. We see that the quantum dynamics are qualitatively different from the semiclassical solutions after some interaction time, an effect which has previously been found in SHG~\cite{tio}, third harmonic generation~\cite{THG2002,THGcascade}, and sum frequency generation~\cite{SFG}. This effect is purely due to the quantum nature of the fields. We note here that the ratio $\kappa_{2}/\kappa_{1}$ has a dramatic effect on the intensities, with $N_{1}$ almost vanishing before its revival when this ratio is set to unity, so that this could be interesting to investigate further.

\begin{figure}[tbhp]
\includegraphics[width=0.75\columnwidth]{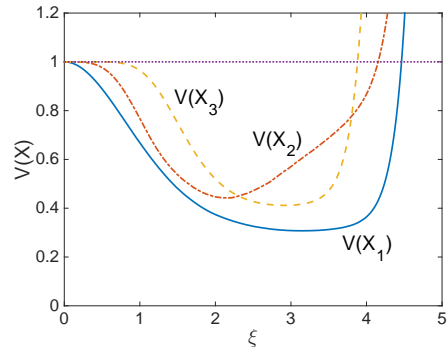}
\caption{(colour online) Positive-P solutions for $\hat{X}$ quadrature variances in the travelling wave configuration, with parameters as in Fig.~\ref{fig:Ntrav}. The line at one is a guide to the eye.}
\label{fig:VXtrav}
\end{figure}

\begin{figure}[tbhp]
\includegraphics[width=0.75\columnwidth]{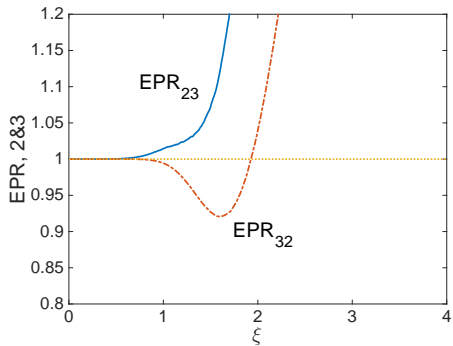}
\caption{(colour online) Positive-P solutions for EPR-steering between $\omega_{2}$ and $\omega_{3}$ in the travelling wave configuration, with parameters as in Fig.~\ref{fig:Ntrav}. The line at one is a guide to the eye.}
\label{fig:EPR23trav}
\end{figure}

For the same parameters as in Fig.~\ref{fig:Ntrav}, we find that there is quadrature squeezing in all three fields during the initial interaction, as shown in Fig.~\ref{fig:VXtrav}. The disappearance of the squeezing after some time is reminiscent of many travelling wave processes~\cite{tio,THG2002,SFG,THGcascade}, and happens once downconversion becomes important, since this is a spontaneous process. In order to detect entanglement, we can use either the Duan-Simon~\cite{Duan,Simon} or the Reid EPR (Einstein-Podolsky-Rosen) criteria~\cite{EPR,EPRMDR}. Since states exhibiting the EPR paradox are a strict subset of the entangled states~\cite{Wisesteer}, we will show results for the Reid EPR criteria, written in terms of products of inferred variances. We label these products $EPR_{jk}$, signifying that mode $j$ can be steered by mode $k$. The two fields which violate the inequalities to the largest degree, i.e. $EPR_{jk}<1$, are the second and fourth harmonics, which exhibit asymmetric EPR-steering~\cite{sapatona,meu,Natexp}, as shown in Fig.~\ref{fig:EPR23trav}. We note here that steering and entanglement are found between other modes, and that the degree of violation of the inequalities depends on the actual parameters, but rather than investigate this further we will move to the intracavity case.

\begin{figure}[tbhp]
\includegraphics[width=0.75\columnwidth]{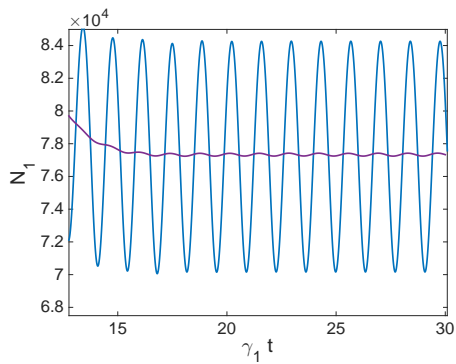}
\caption{(colour online) Positive-P and classical solutions for $N_{1}$ in the self-pulsing regime, for $\kappa_{1}=0.005$, $\kappa_{2}=4\kappa_{1}$, $\gamma_{1}=1$, $\gamma_{2}=\gamma_{3}=\gamma_{1}/2$, and $\epsilon=400$. The classical solution has the larger oscillations.}
\label{fig:pulse}
\end{figure}

In the intracavity configuration we find that the semi-classical and quantum solutions for the intensities are identical until a certain pump power, after which the system enters a self-pulsing regime~\cite{Yerevan,SHGpulse}.
We find that, much like the case of intracavity third harmonic generation~\cite{KVK3,THGcascade}, the classical solutions overstate the amplitude of the pulsing. This is shown in Fig.~\ref{fig:pulse} and does not happen in SHG, where the quantum and semi-classical solutions are identical. This is presumably because there is a clear hard mode transition in SHG, where two eigenvalues of the drift matrix develop imaginary and conjugate eigenvalues. The eigenvalue spectra of this cascaded system is more complicated, and we were not able to find analytical solutions. However, we have ensured numerically that the spectra given below are in a stable regime. 

When nonlinear optical media are held inside a pumped optical cavity, the measured observables are usually the output spectral correlations, which are accessible using homodyne measurement techniques~\cite{mjc}. These are readily calculated in the steady-state by treating the system as an Ornstein-Uhlenbeck process~\cite{SMCrispin}. In order to do this, we begin by expanding the positive-P variables into their steady-state expectation values plus delta-correlated Gaussian fluctuation terms, e.g.
\begin{equation}
\alpha_{ss} \rightarrow \langle\hat{a}\rangle_{ss}+\delta\alpha.
\label{eq:fluctuate}
\end{equation}
Given that we can calculate the $\langle\hat{a}\rangle_{ss}$, we may then write the equations of motion for the fluctuation terms. The resulting equations are written for the vector of fluctuation terms as
\begin{equation}
\frac{d}{dt}\delta\vec{\alpha} = -A\delta\vec{\alpha}+Bd\vec{W},
\label{eq:OEeqn}
\end{equation}
where $A$ is the drift matrix containing the steady-state solution, $B$ is found from the factorisation of the drift matrix of the original Fokker-Planck equation, $D=BB^{T}$, with the steady-state values substituted in, and $d\vec{W}$ is a vector of Wiener increments. As long as the matrix $A$ has no eigenvalues with negative real parts, this method may be used to calculate the intracavity spectra via
\begin{equation}
S(\omega) = (A+i\omega)^{-1}D(A^{\mbox{\small{T}}}-i\omega)^{-1},
\label{eq:Sout}
\end{equation}
from which the output spectra are calculated using the standard input-output relations~\cite{mjc}.

In this case
\begin{equation}
A =
\begin{bmatrix}
\gamma_{1} & -\kappa_{1}\alpha_{2} & -\kappa_{1}\alpha_{1}^{\ast} & 0 & 0 & 0 \\
-\kappa_{1}\alpha_{2}^{\ast} & \gamma_{1} & 0 & -\kappa_{1}\alpha_{1} & 0 & 0 \\
\kappa_{1}\alpha_{1} & 0 & \gamma_{2} & -\kappa_{2}\alpha_{3} & -\kappa_{2}\alpha_{2}^{\ast} & 0 \\
0 & \kappa_{1}\alpha_{1}^{\ast} & -\kappa_{2}\alpha_{3}^{\ast} & \gamma_{2} & 0 & -\kappa_{2}\alpha_{2} \\
0 & 0 & \kappa_{2}\alpha_{2} & 0 & \gamma_{3} & 0 \\
0 & 0 & 0 & \kappa_{2}\alpha_{2}^{\ast} & 0 & \gamma_{3}
\end{bmatrix},
\label{eq:Amat}
\end{equation}
and $D$ is a $6\times 6 $ matrix with $\left[\kappa_{1}\alpha_{2},\kappa_{1}\alpha_{2}^{\ast},\kappa_{2}\alpha_{3},\kappa_{2}\alpha_{3}^{\ast},0,0\right]$ on the diagonal. In the above, the $\alpha_{j}$ should be read as their steady-state values.
Because we have parametrised our system using $\gamma_{1}=1$, the frequency $\omega$ is in units of $\gamma_{1}$. $S(\omega)$ then gives us products such as $\delta\alpha_{i}\delta\alpha_{j}$ and  $\delta\alpha_{i}^{\ast}\delta\alpha_{j}^{\ast}$, from which we construct the output variances and covariances for modes $i$ and $j$ as
\begin{equation}
S^{out}(X_{i},X_{j}) = \delta_{ij}+\sqrt{\gamma_{i}\gamma_{j}} \left(S_{ij}+S_{ji}\right).
\label{eq:Sout}
\end{equation}

\begin{figure}[tbhp]
\includegraphics[width=0.75\columnwidth]{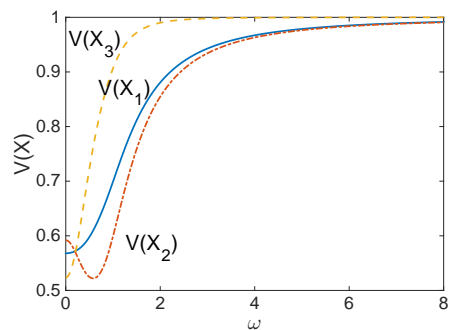}
\caption{(colour online) $\hat{X}$ quadrature variances, for $\kappa_{1}=0.005$, $\kappa_{2}=4\kappa_{1}$, $\gamma_{1}=1$, $\gamma_{2}=\gamma_{3}=\gamma_{1}/2$, and $\epsilon=105$. The frequency axis is in units of the linewidth of the fundamental, $\gamma_{1}$.}
\label{fig:VXcav}
\end{figure}

\begin{figure}[tbhp]
\includegraphics[width=0.75\columnwidth]{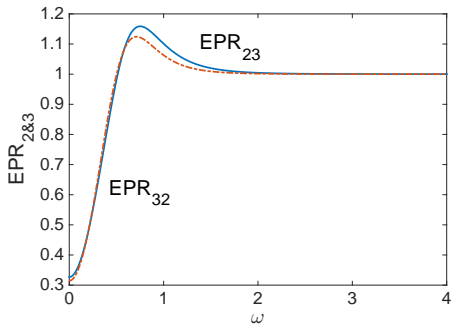}
\caption{(colour online) The Reid criteria for steering between $\omega_{2}$ and $\omega_{3}$, for the same parameters as Fig.~\ref{fig:VXcav}.}
\label{fig:EPR23}
\end{figure}

The first property we examine here is again the quadrature variances. As shown in Fig.~\ref{fig:VXcav}, a degree of squeezing comparable to that found in SHG~\cite{MJCDFW} is available in all three modes. We also find that EPR-steering exists for all of the three possible bipartitions. That which violates the inequality by the greatest amount is the grouping of mode two with mode three, as shown in Fig.~\ref{fig:EPR23}. This violation necessarily also means that these two modes are entangled, across one octave of frequency difference. Non-classical correlations have previously been predicted over such a frequency ratio~\cite{sumdiff}, while  entanglement and EPR steering have been predicted in SHG in both travelling wave~\cite{UFFdia} and intracavity configurations~\cite{meu}, and experimentally achieved via an intracavity $\chi^{(2)}$ medium pumped at both the fundamental and harmonic frequencies~\cite{Grosse}. Depending on the pump frequency, the steering found here could extend from the optical to the ultraviolet and be useful for multiplexing in quantum communication applications~\cite{Baune}.

\begin{figure}[tbhp]
\includegraphics[width=0.75\columnwidth]{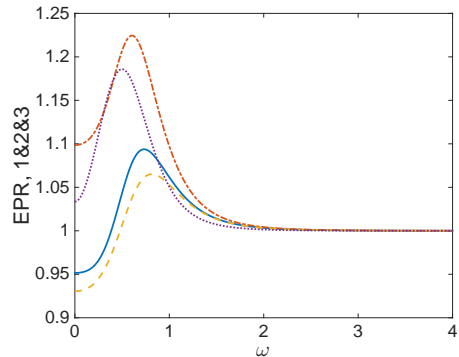}
\caption{(colour online) The Reid criteria for steering between modes one and two, and one and three, for the same parameters as Fig.~\ref{fig:VXcav}. The solid line is $EPR_{12}$, the dash-dotted line is $EPR_{21}$, the dashed line is $EPR_{13}$, and the dotted line is $EPR_{31}$.}
\label{fig:EPR123}
\end{figure}

When we consider the fundamental and its entanglement with the second and fourth harmonics, we again find that entanglement is present for both bipartitions, as shown by the EPR steering results in Fig.~\ref{fig:EPR123}. In this case we find that the steering in both bipartitions is asymmetric, with the fundamental being able to steer the higher modes but these being unable to steer the fundamental. Since this system is Gaussian, the Reid criteria are both necessary and sufficient to show this feature.

In conclusion, we have proposed and analysed a cascaded $\chi^{(2)}$ system which has outputs that are bipartite entangled and EPR states over a two octave frequency range. With an input at the $1550$nm frequency, for example, this could produce entangled modes in both the infrared and visible ranges. With a $1064$nm input, the output frequencies range from infrared through the visible, to ultraviolet. This frequency range lends itself to multiplexing applications. The existence of asymmetric EPR steering between the fundamental and the higher modes makes this device even more versatile, with possible uses in quantum key distribution and continuous variable teleportation.



\end{document}